\documentclass[aps,pra,reprint,amsmath,amssymb,floatfix]{revtex4-1}

\usepackage{bm,graphicx,amsmath,amssymb,gensymb,siunitx}
\usepackage{hyperref}

\hypersetup{
	colorlinks=true,
	linkcolor=blue,
	citecolor=blue,
	filecolor=magenta,
	urlcolor=cyan
}

\graphicspath{
{}
{figures/}
{../figures/}
}

\allowdisplaybreaks[1]

\newcommand{\la}{\langle}
\newcommand{\ra}{\rangle}
\newcommand{\kk}{{\bm{k}}}

\newcommand{\rr}{\bm{r}}

\newcommand{\nn}{{\bm{n}}}

\newcommand{\ttau}{\bm{\tau}}

\newcommand{\ssigma}{\bm{\sigma}}

\newcommand{\hnn}{\hat{\nn}}
\newcommand{\htt}{\hat{\bm{t}}}
\newcommand{\hzz}{\hat{\bm{z}}}

\newcommand{\e}{\text{e}}
\newcommand{\ii}{\text{i}}

\newcommand{\cR}{\mathcal{R}}
\newcommand{\cM}{\mathcal{M}}
\newcommand{\dtheta}{\Delta\theta}
\newcommand{\cH}{\mathcal{H}}
\newcommand{\dH}{\Delta_\text{H}} 
\newcommand{\dS}{\Delta_\text{S}}

\newcommand{\papertitle}{Kane-Mele with a twist: Quasicrystalline higher-order topological insulators with fractional mass kinks}

\newcommand{\tcm}{T.C.M. Group, Cavendish Laboratory, University of Cambridge, JJ Thomson Avenue, Cambridge, CB3 0HE, U.K.}

\begin{document}
	
\title{\papertitle}

\author{Stephen Spurrier}
\author{Nigel R. Cooper}

\affiliation{\tcm}

\date{\today}


\begin{abstract}
	We establish an analytic low-energy theory describing higher-order topological insulator (HOTI) phases in quasicrystalline systems. We apply this to a model consisting of two stacked Haldane models with oppositely propagating edge modes, analogous to the Kane-Mele model, and with a \ang{30} twist. We show that the resulting localized modes at corners, characteristic of a HOTI, are not associated with conventional mass inversions but are instead associated with what we dub ``fractional mass kinks''. By generalizing the low-energy theory, we establish a classification for arbitrary $ n $-fold rotational symmetries. We also derive a relationship between corner modes in a bilayer and disclination modes in a single layer. By using numerics to go beyond the weak-coupling limit, we show that a hierarchy of additional gaps occurs due to the quasiperiodicity, which also harbor corner-localized modes. 
\end{abstract}

\maketitle


\section{Introduction}

The topological classification of insulators with internal symmetries is a cornerstone of modern condensed-matter theory~\cite{kruthoff2017topological,kitaev2009periodic,ryu2010topological,qi2011topological,chiu2016classification}. Recently, this classification was further enriched by the addition of crystalline symmetries~\cite{fu2011topological,fang2012bulk,slager2013space,ando2015topological}. In fact, a full classification has been achieved for all 230 crystal symmetry groups without internal symmetries~\cite{po2017symmetry,bradlyn2017topological,vergniory2019complete}. A particularly interesting result in this direction is the discovery of higher-order topological insulator (HOTI) phases~\cite{parameswaran2017topological,fang2017rotation,benalcazar2017quantized,benalcazar2017electric,song2017dimensional,langbehn2017reflection,schindler2018higher}. In these, crystalline symmetries allow for a generalized bulk-boundary correspondence where the insulating $ D $-dimensional bulk has $ (D-d) $-dimensional edge states with $ d>1 $.

Another fascinating line of research for topological insulators in recent years has been in exploring their relevance for quasicrystals~\cite{kraus12topologicalstates,tran2015topological,bandres16topological,fulga2016aperiodic}. These are systems that are intermediate between periodic and amorphous systems, in that they possess long-range order and yet lack translational symmetry~\cite{shechtman84metallic,steinhardt1987physics,trebin2006quasicrystals}. They are defined as having Bragg peaks (and hence long-range order) that require more basis vectors than spatial dimensions to index (therefore lacking a Brillouin zone)~\cite{steinhardt1987physics}. For topological insulators, quasicrystals provide a test bed for exploring the extent to which these are robust to disorder~\cite{tran2015topological,bandres16topological,fulga2016aperiodic}. Through the Harper-Hofstadter model~\cite{harper1955single,hofstadter76energy}, these have also been shown to possess topological indices that are inherited from higher-dimensional parent systems via a projection, suggesting routes to studying higher-dimensional topology experimentally~\cite{kraus12topologicalstates,kraus13fourdimensional,dana2014topologically,ozawa2019topological}.

For higher-order topological insulators, quasicrystals present a particularly interesting prospect. Since they do not possess translational symmetry, all known classifications based on symmetry indicators in the Brillouin zone no longer apply~\cite{khalaf2018symmetry,benalcazar2019quantisation}. While this is also true for amorphous systems, which have been shown to support HOTI corner modes~\cite{agarwala2020higherorder}, quasicrystals notably still possess rotational symmetries. Moreover, as these rotational symmetries can be disallowed for crystalline systems, and are absent in amorphous systems, a HOTI phase here would be unique to a quasicrystal. It was shown in recent studies that quasicrystals do indeed support HOTI phases~\cite{chen2019higher}, including phases with disallowed rotational symmetry~\cite{varjas2019topological}. All current approaches rely solely on numerical methods, lacking a direct analytical understanding.

Here we bridge this gap by constructing an analytical approach that can describe quasicrystalline HOTIs. We demonstrate this approach for a simple model consisting of two Haldane models~\cite{haldane88model} stacked with a \ang{30} twist, which we liken to Kane-Mele~\cite{kane2005quantum,kane2005z2} with a twist. This model is, by construction, quasicrystalline since it is two-dimensional and has Bragg peaks indexed by four basis vectors~\cite{steinhardt1987physics}. Our analytical approach is based on a low-energy edge theory. Unlike the Kane-Mele model in which the edge theory is protected by a local time-reversal symmetry, the nonlocal 12-fold rotational symmetry in the model we study does not protect the edge modes from gapping out. Instead, the rotational symmetry places a constraint on a phase $ \theta $ parametrizing the edge mass. This forces domain walls at the corners, resulting in the corner modes associated with the HOTI phase.

Interestingly, the domain walls in the mass are not the standard ``mass inversions'' (corresponding to $ \dtheta = \pi $) with the associated corner charge of $ 1/2 $, encountered in similar studies~\cite{fang2017rotation,khalaf2018symmetry}. Instead, for the model we study, these involve a fractional phase shift, $ \dtheta = \pi/2 $, corresponding to a corner charge of $ 1/4 $~\cite{wang2019boundary}. We therefore dub this a ``fractional mass kink''. Moreover, we generalize this result to arbitrary $ n $-fold rotational symmetry, finding that corner charges are fractionalized as $ Q = p/n $ associated with a fractional mass kink $ \dtheta = 2\pi p/n $, where $ p $ is an integer. Interestingly, this provides an alternative perspective on a classification of $ C_n $-protected corner charges in Ref.~\cite{benalcazar2019quantisation} and generalizes to arbitrary rotational symmetries, including quasicrystalline.

Furthermore, in Sec.~\ref{sec:disclinations}, we use our low-energy theory to make a connection between the corner modes of a bilayer and disclination modes in a single layer. In doing so, we extend the known disclination modes of a single Haldane layer~\cite{ruegg2013bound} to a generalized relationship between disclination charge and disclination angle applicable to arbitrary rotational symmetries. In Sec.~\ref{sec:hierarchy}, we use numerics to go beyond the low-energy theory (weak-coupling limit) and find that at stronger couplings a hierarchy of gaps opens in the edge spectrum, with these harboring additional corner-localized modes. We show that this is a direct result of the quasiperiodicity and therefore provides a striking example of how quasicrystalline HOTIs differ from their crystalline counterparts. In Sec.~\ref{sec:generalisations}, we discuss further generalizations and highlight an interesting feature of the twist construction. We show that a trivial HOTI without a twist can be nontrivial after twisting.

\section{Model}

\begin{figure}[tbp]
	\centering
	\includegraphics[trim={.0cm .0cm 0.cm .0cm}, clip,width=.99\linewidth]{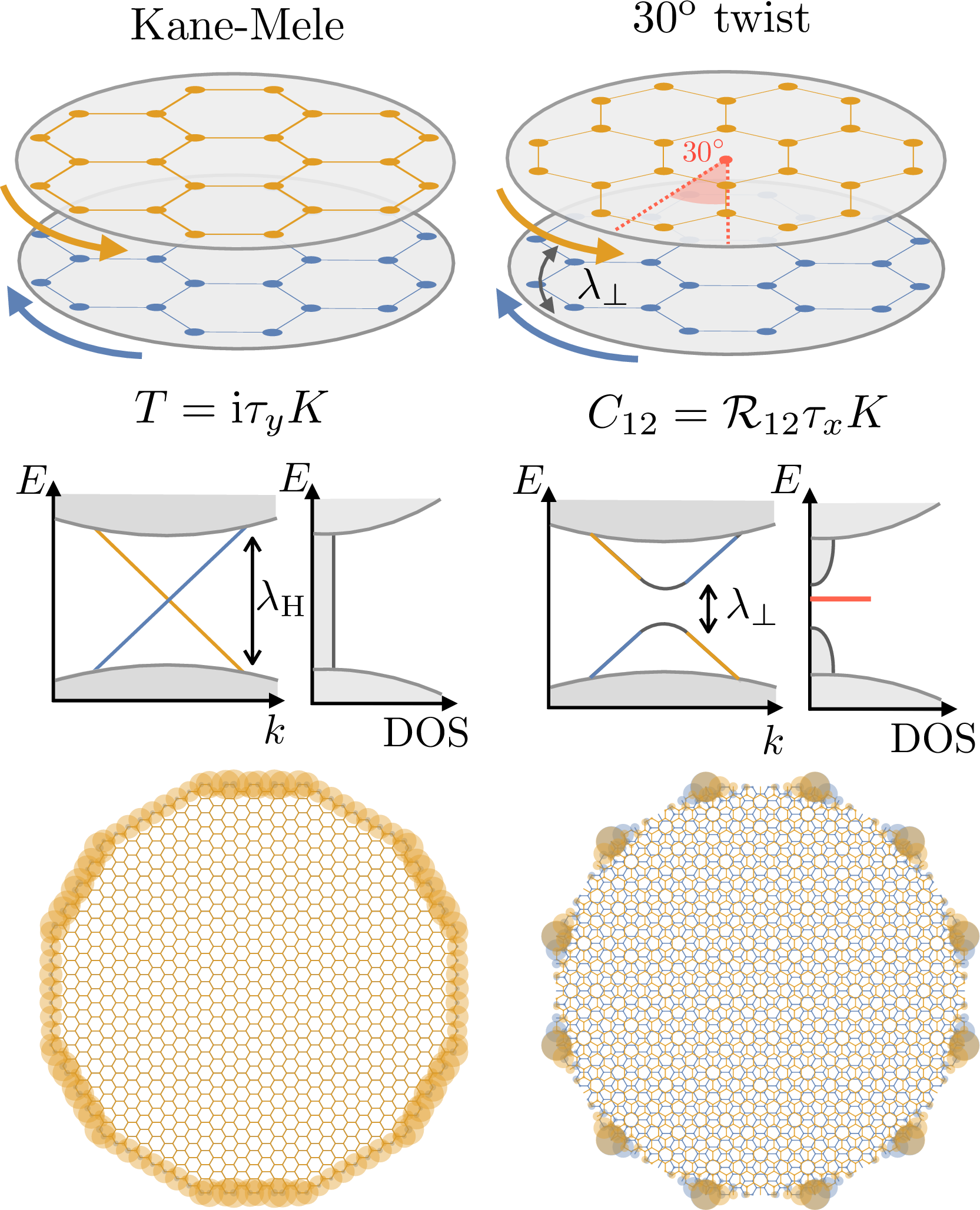}%
	\caption{\label{fig:C12HOTI}
		Left: The Kane-Mele model amounts to stacking two Haldane models with opposite Chern numbers and requiring a spinful time-reversal symmetry $ T = \ii \tau_y K $. As a result of this symmetry, the two edge modes are protected from gapping out. Right: In the model we study, we imagine twisting one of the Haldane layers by \ang{30}. As such, we remove the \emph{local} time-reversal symmetry and replace this with a \emph{nonlocal} $ 12 $-fold rotation plus time-reversal symmetry. Without the local time-reversal symmetry, the edge modes are gapped. However, due to the rotational symmetry the mass that gaps the edge modes changes by a phase between edges, protecting corner-localized modes.
	}
\end{figure}

We study a model that is constructed by stacking two Haldane models~\cite{haldane88model} with opposite Chern numbers with a \ang{30} twist, as shown in Fig.~\ref{fig:C12HOTI}. It is given by
\begin{align}
H = -t \sum_{\la i j \ra} c_i^\dagger \tau_0 c_j + \lambda_\text{H} \sum_{\la \la i j \ra \ra}  \ii \nu_{ij} c_i^\dagger \tau_z c_j + \lambda_\perp \sum_{ij} t_{ij}^\perp c_i^\dagger \tau_x c_j, \label{eq:tightbinding}
\end{align}
where $ c_i = \begin{pmatrix} c_i^\text{t} & c_i^\text{b} \end{pmatrix}^\text{T} $ is a two-component spinor with components acting on the top and bottom layers described by the Pauli matrices $ \tau_x $, $ \tau_y $, $ \tau_z $ and the identity $ \tau_0 $. The annihilation operators $ c_i^\text{t} $ ($ c_i^\text{b} $) act on the site at $ \rr_i^\text{t} $ ($ \rr_i^\text{b}) $, with $ \rr_i^\text{t} = \cR_{12} \rr_i^\text{b} + d_z \hzz $, where $ \cR_n $ is a rotation by $ 2\pi/n $, and $ d_z $ is the interlayer separation. The nearest- and next-nearest-neighbor notation $ \la \dots \ra $ and $ \la\la \dots \ra\ra $ denotes a coupling between sites on the same layer. 
The first two terms describe the  Haldane models on each layer, while the third is an all-to-all interlayer hopping governed by $  t_{ij}^\perp = \exp \left( -|\rr_i-\rr_j|/\delta \right) $, where top and bottom labels have been suppressed. 

In Fig.~\ref{fig:C12HOTI} we outline a natural comparison of the model we study to that of Kane-Mele~\cite{kane2005z2,kane2005quantum}. Indeed, by associating the spin degree of freedom there with a physical layer degree of freedom here, giving these layers a \ang{30} twist and replacing the ``Rashba'' interlayer coupling with that in  \eqref{eq:tightbinding}, one arrives at the model studied here. The crucial difference, however, is in the relevant symmetry. For Kane-Mele, the symmetry $ \ii \tau_y K $ is a \emph{local} symmetry, in the sense that it relates spatially local degrees of freedom (albeit applied globally over the sample), whereas here the relevant symmetry, $ R_{12}\tau_x K $,  is \emph{nonlocal}, relating spatially separated degrees of freedom. The local symmetry in the Kane-Mele model is sufficient to protect the two oppositely propagating edge states from being gapped out everywhere along the edge, providing an example of a spinful time-reversal symmetry-protected topological insulator. However, the nonlocal symmetry here does not protect the edge states from being gapped out. Instead, this nonlocal crystalline symmetry protects lower-dimensional corner modes at the intersections of the 12 edges of a 12-fold rotationally symmetric sample. Our model therefore provides an example of a higher-order topological insulator protected by a crystallographically disallowed 12-fold rotational symmetry.

\begin{figure}[tbp]
	\centering
	\includegraphics[trim={.0cm 0.0cm .0cm .0cm}, clip,width=.75\linewidth]{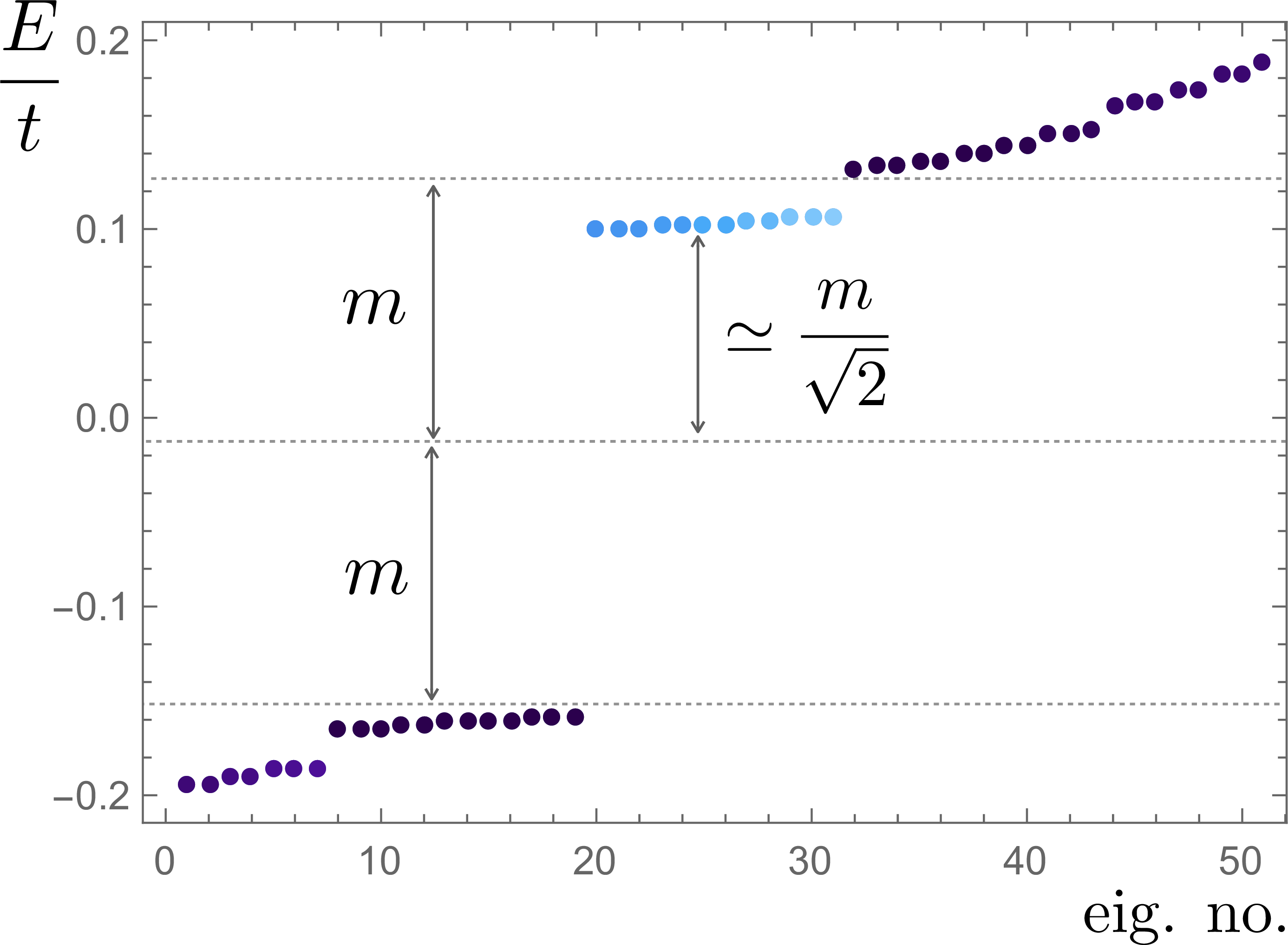}%
	\caption{\label{fig:msqrt2}
		A close-up of the numerically obtained spectrum of \eqref{eq:tightbinding} with parameters $ \lambda_\text{H} = 0.3t $, $ \lambda_\perp = 3.0t $, $ d_z = 1.362 a  $, and $ \delta = 0.184a $. The 12 corner-localized states at $ E/t \simeq 0.1 $ are indicated by light-blue coloring. Unlike corner modes due to a conventional mass inversion ($ \dtheta = \pi $ in our notation) in which the localized modes are at $ E=0 $, the mass here undergoes a fractional mass kink ($ \dtheta = \pi/2 $), resulting in a bound-state energy of $ E = m/\sqrt{2} $, where $ m $ is the half-gap width. Due to finite-size effects, the degeneracy between the 12 corner modes is lifted; this splitting reduces exponentially with system size.
	}
\end{figure} 

We demonstrate numerically the presence of a higher-order topological phase with crystallographically disallowed rotational symmetry by directly computing the spectrum, with the results shown in Fig.~\ref{fig:C12HOTI}. For the parameters $ t=1 $, $ \lambda_\text{H}  = 0.3 $, and $ \lambda_\perp = 1 $, one finds a gap opens in the edge spectrum, as outlined schematically in Fig.~\ref{fig:C12HOTI}. In Fig.~\ref{fig:msqrt2}, we show a close-up of the spectrum at this gap opening. One sees 12 in-gap corner-localized states, with these indicated by a light-blue coloring. An example of a corner-localized eigenstate is shown in Fig.~\ref{fig:C12HOTI}.

Diagnosing this nontrivial higher-order topology is hindered by the presence of the quasiperiodicity. Established tools based on eigenvalues at high-symmetry points in the Brillouin zone are ruled out as momentum is no longer a good quantum number. Instead, in the following we pursue an approach based on classifying how crystalline symmetries can enforce domain walls in a low-energy edge theory. As this approach does not rely on crystal momentum being a good quantum number, it is robust in the presence of quasiperiodicity.

\section{Low-energy theory and classification\label{sec:lowenergy}}

In order to understand the origin of the HOTI phase we ask how the rotational symmetry of the model we study constrains the low-energy theory at the edge. We construct the low-energy theory using the following arguments. Considering the $ \lambda_\perp = 0 $ limit, our model reduces to two uncoupled Haldane models of opposite chirality, and therefore, we expect a gapless 1D Dirac theory consisting of a single term, $ k \tau_z $, describing the two counterpropagating modes on the edge (see Appendix~\ref{app:derivation}). For $ \lambda_\perp \neq 0 $, a gap is opened in the edge spectrum, as shown in Fig.~\ref{fig:msqrt2}. To describe this, the low-energy theory must include additional terms that do not commute with $ k\tau_z $. We therefore include terms proportional to $ \tau_x  $ and $ \tau_y $. Including these alongside the kinetic term, one has the following low-energy theory:
\begin{align}
\mathcal{H} = k \tau_z + m\mathcal{M}_\theta,\label{eq:lowenergy}
\end{align}
with $ \mathcal{M}_\theta \equiv  \cos \theta\, \tau_x +  \sin \theta\, \tau_y $. The mass terms $ \tau_x $, $ \tau_y $ have been parametrized via $ m $ and $ \theta $, with both parameters considered to be  functions of position along the edge. 

\begin{figure}[tbp]
	\centering
	\includegraphics[trim={0.cm 0.cm 0.cm .0cm}, clip,width=.99\linewidth]{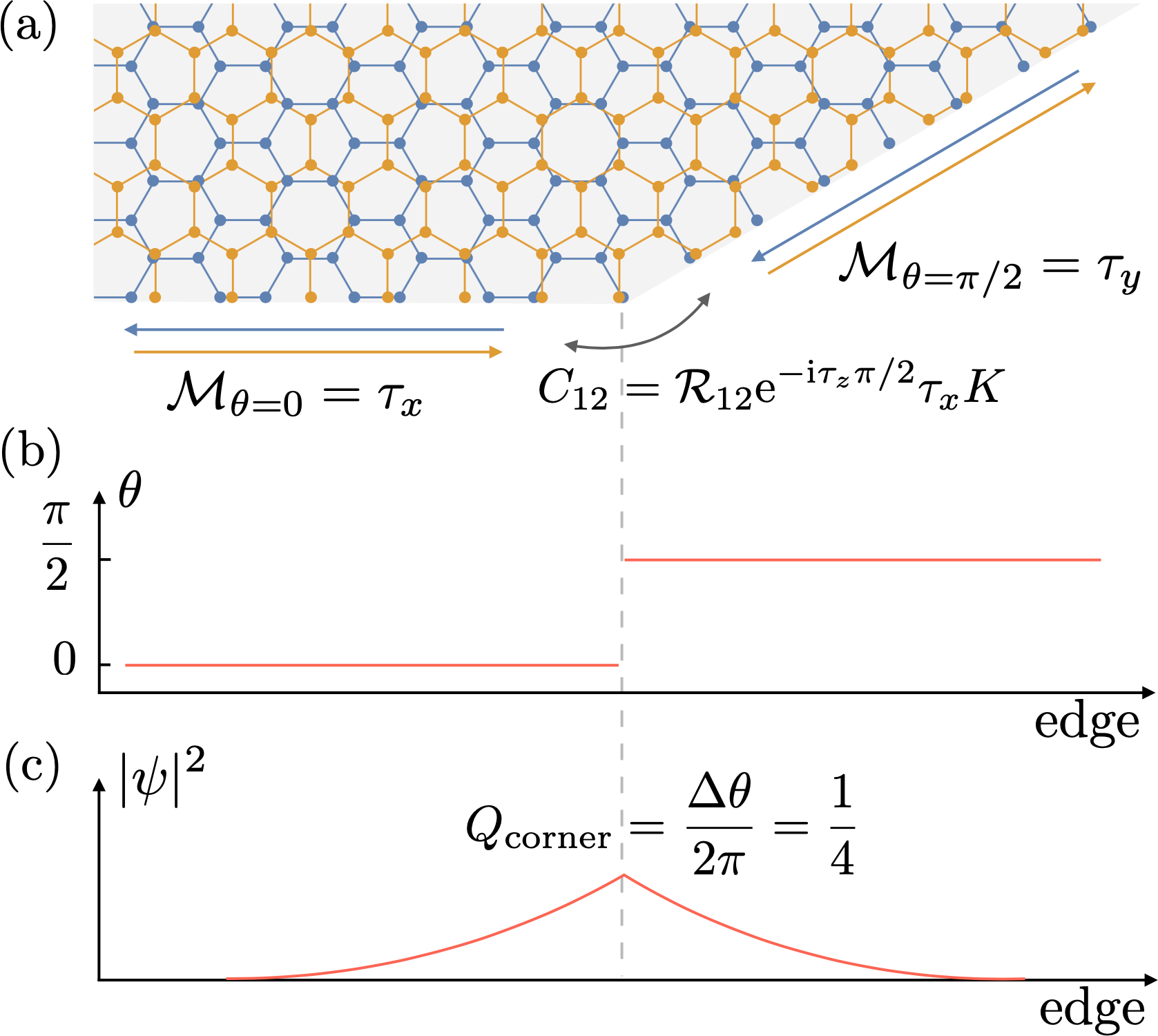}%
	\caption{\label{fig:fracmass}
		(a) The corner-localized modes are understood in a low-energy theory consisting of left- and right-propagating modes coupled by a mass $ \cM_\theta $. (b) The effective $ C_{12} $ symmetry in the low-energy theory forces the parameter $ \theta $ to change by $ \pi/2 $ at a corner of a dodecagonal sample. (c) At this domain wall in $ \theta $, there exists a localized state with a fractional fermion charge of $ 1/4 $. This is the corner mode found numerically.
	}
\end{figure}

If one had $ m=0 $ along the entire edge, the edge spectrum would be gapless. This is the case for the Kane-Mele model, as there the local spinful time-reversal symmetry $ \ii \tau_y K $ anticommutes with $ \tau_x $ and $ \tau_y $ and therefore forces $ m=0 $ everywhere along each edge. However, here we will show in the following that the nonlocal $ C_{12} $ symmetry of the bilayer model imposes a constraint that forces a kink in the angle $ \theta $ across a corner, as shown in Fig.~\ref{fig:fracmass}.

Before discussing the specifics of the low-energy theory of the model we study, we first outline a general classification for how rotational symmetries can lead to HOTI phases for the low-energy edge theory in \eqref{eq:lowenergy}. Consider all symmetries $ C_n $ acting on the edge theory that contain an $ n $-fold rotation $ \cR_n $ alongside an additional unitary or antiunitary operation $ U_n $, that is, $ C_n = \cR_n U_n  $. If the operator $U_n$ is antiunitary it will contain a complex conjugation $ K $, and the associated $ C_n $ will amount to a rotation plus time reversal. The only conditions we require for $ C_n $ are that a full rotation is spinor-like, $ C_n^n = -1 $~\cite{spinorlike}, and that $ C_n $ commutes with the kinetic part of the low-energy theory $ \left[ C_n , k\tau_z \right] = 0 $. This constrains $ C_n $ to the following representations. For unitary $ U_n $ one has $ C_n = \cR_n \exp[ -\ii \pi (q \tau_0 + p \tau_z)/n ]$, with integer $ q $ and $ p $, with $ q + p $ being odd.
However, for antiunitary $ U_n $, one has $ C_n = \cR_n \exp\left(-\ii \tau_z \pi p/n \right)\tau_x K $ for even $ n $ and odd $ p $. The requirement of even $ n $ in the antiunitary case is because only even powers of an antiunitary operator are unitary.

In order for the mass $ m\cM_\theta $ to be present it must commute with $ C_n $, that is, $ \left[C_n, m\cM_\theta \right] =0 $. As $ C_n $ acts nonlocally on the edge, this condition relates masses on neighboring edges via
\begin{align}
U_n \cM_\theta U_n^\dagger = \cM_{\theta+2\pi p /n},
\end{align}
for both unitary and antiunitary representations. That is, $ C_n $ causes rotations in the phase $ \theta $ between two edges. As is well known from the work of Jackiw and Rebbi~\cite{jackiw1976solitons}, a phase shift of $ \pi $, known as a mass inversion, results in a zero-energy localized state with fractional charge, $ Q=1/2 $. While this can be the case here, more generally, the phase shift $ \dtheta = 2\pi p /n $ between two edges is a multiple of $ 2\pi/n $. Nevertheless, for any nonzero $ \dtheta $ the following localized state can be shown to exist~\cite{wang2019boundary}:
\begin{align}
\psi(x) = \frac{1}{\sqrt{2}}\begin{pmatrix}
\e^{-\ii(\theta+\dtheta/2)}\\
1
\end{pmatrix}
\e^{-\kappa |x|},
\end{align}
where $ \kappa = m \sin\left(\dtheta/2\right) $ and $ x $ is the distance from the domain wall. One can also show that this localized state lies at an energy of $ E = m \cos\left(\dtheta/2\right) $ and has a quantized charge of $ Q = \dtheta/2\pi = p/n $~\cite{goldstone1981fractional,jackiw1983continuum,qi2008fractional}. Therefore, the conventional mass inversion, $ p/n = 1/2 $, is a special case. More generally, rotational symmetries bind fractional charges that are multiples of $ 1/n $.

A few comments can be made about these results. The trivial case for which there is no domain wall, $ \dtheta=0 $, has $ \kappa = 0 $, $ E = m $, and $ Q = 0 $. That is, this state is delocalized (infinite localisation length), it is part of the edge band, and it has zero quantized charge. Therefore, for this representation ($ p=0 $), the state is trivial. Note that as $ p=0 $ is not possible for antiunitary representations, all antiunitary representations are nontrivial. 
For $ \dtheta = \pi $, that is, representations with $ p/n = 1/2 $, one recovers the familiar case of a mass inversion. Here one finds a maximally localized state, $ \kappa = m $, with zero energy, $ E = 0 $, and half fermion charge, $ Q = 1/2 $. For all other representations one finds a fractional mass kink. These have nonzero energy and a fractional charge that is a multiple of $ 1/n $ different from $ 1/2 $.

We highlight two crucial features of this classification. The first is that our theory applies to all rotational symmetries regardless of whether these are allowed by a periodic lattice. As such, our classification applies to quasicrystalline systems, which are typically not included in HOTI classifications due to the lack of a well-defined Brillouin zone.
Another interesting feature is the presence of fractional mass kinks. In three-dimensional (3D) class-AII classifications, fractional mass kinks are not possible~\cite{khalaf2018symmetry}, which can be understood from the following. The two-dimensional (2D) edge theory (of the 3D TI) has two Pauli matrices assigned to the two momentum components, leaving a single Pauli matrix for the mass. Therefore, symmetries cannot ``rotate'' the mass as in one-dimension but can still invert the sign of the mass. An interesting consequence of this is that odd rotational symmetries are necessarily trivial in three-dimensions; however, in 2D these can be nontrivial.

\begin{figure}[tbp]
	\centering
	\includegraphics[trim={.0cm 0cm 0.0cm .0cm}, clip,width=.99\linewidth]{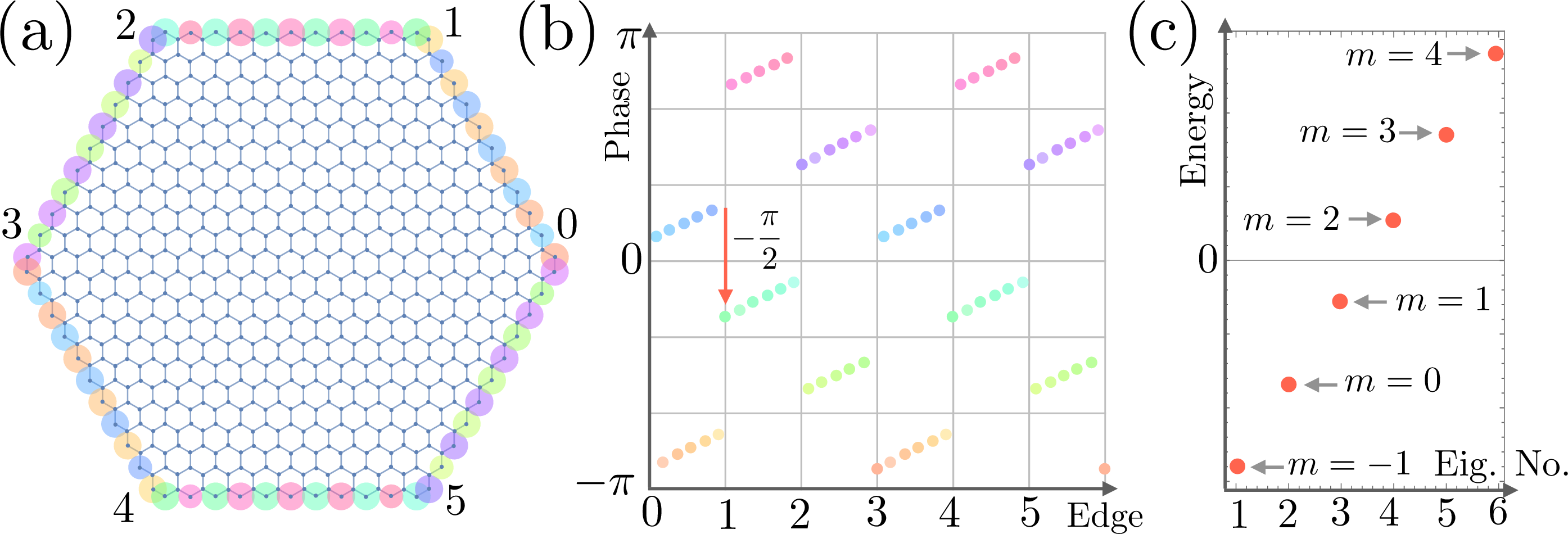}%
	\caption{\label{fig:C6phases} (a) and (b) Plots of the phase winding of a Haldane model edge state. One observes $ \pi/2 $ shifts at the corners of a hexagonal sample, indicating that the effective sixfold rotational symmetry for the low-energy theory is $ C_6 = \pm\ii\cR_6 $. (c) The same $ \pi/2 $ phase shift is also revealed by an overall shift in the numerically obtained low-energy $ m=0 $ angular momentum edge state.
	}
\end{figure}

We also highlight that unlike chiral or particle-hole symmetry-protected phases, in which the symmetry protects the precise location of the zero-dimensional state within the gap, the rotational symmetry does not exclude local terms that can move the energy of the corner states~\cite{peterson2020fractional}. Instead, the rotational symmetry protects the corner \emph{charge}. In the low-energy theory there are no symmetry-allowed terms that can remove the twist in the mass. Since it is the kinks in this mass that determine the charge of the corner state, it is the charge that is symmetry protected. Furthermore, since the representation of the rotation at the edge can be changed only by closing the bulk gap, the presence of these symmetry-protected charges is the signature of a topologically nontrivial bulk phase.

In order to apply the general theory above to the model we study we need to determine the representation for the $ C_{12} $ edge symmetry. To do this, we turn to the simpler question of finding the correct representation for the $ C_6 $ symmetry of a single layer. We discuss how this can be determined and then show how the full $ C_{12} $ symmetry can be obtained by taking a ``square root'' of the $ C_6 $ symmetry.

Consider the phase of a low-energy chiral edge state on a hexagonal sample, as shown in Fig.~\ref{fig:C6phases}. Away from the corners, one finds a phase winding, $ \exp\left(\ii k x\right) $, corresponding to a state with momentum $ k = \pi/a + \delta k $, where $ a $ is the 1D unit cell length. That is, the phase shifts by $ \pi $ between neighboring unit cells, which is expected since the chiral edge state for a ``zig-zag'' edge passes through $ k=\pi/a $. This is accompanied by a small linear increase due to the small but nonzero energy of the state. However, at the corners, one sees an abrupt shift of $ \pm\pi/2 $, with the sign determined by the chirality of the edge state. This result is also consistent with studies on disclination modes in the Haldane model~\cite{ruegg2013bound,wang2019boundary,liu2019shift}. This shows that across the corner the state acquires a factor of $ \pm \ii $. Consequently a bilayer will have  $ C_6 = \ii \tau_z = \exp\left(\ii\tau_z 3\pi /6\right) $, and therefore, the $ C_6 $ representation has $ p=3 $.

This result can be corroborated by studying the details of the low-energy spectrum of this finite hexagonal sample, shown in Fig.~\ref{fig:C6phases}(c). Since the system has sixfold rotational symmetry, each eigenstate $ \psi $ can be assigned an angular momentum quantum number $ m $ according to $ \cR_6 \psi = \e^{\ii m \pi/3} \psi $. Given an edge state with angular momentum $ m $, the total phase accumulated along an edge of length $ L $, consisting of the kinetic term $ k x $ plus possible phase shifts $ \varphi $ at a corner, is constrained by $ kL + \varphi = m\pi/3 $. Since the edge state dispersion is $ \epsilon = v_\text{F} k $, one has the following low-energy spectrum of states:
\begin{align}
E_m = \frac{\pi v_\text{F}}{3 L}\left(m - \frac{6\varphi}{2\pi}\right). \label{eq:edgequantisation}
\end{align}
By comparison to numerics, one finds that the $ m = 0 $ state is shifted by $ -3/2 $ of the energy spacing. Therefore, one has $ 6\varphi/2\pi = 3/2 $, which gives $ \varphi = \pi/2 $, in agreement with the previous discussion.

We deduce the full $ C_{12} $ symmetry from the $ C_6 $ symmetry by taking a square root. We require $ C_{12}^2 = C_6 $ and, additionally, look for solutions that include time reversal since we expect the effective edge $ C_{12} $ symmetry to be of a form similar to the bulk $ C_{12} $ symmetry. According to our classification, this amounts to keeping $ p=3 $ fixed while doubling $ n $, giving $ C_{12} = \exp(-\ii \tau_z \pi/4)\tau_x K $. The effect of this symmetry is outlined in Fig.~\ref{fig:fracmass}. The $ C_{12} $ symmetry induces fractional mass kinks, $ \dtheta = \pi/2 $, localizing states at the corners, with energy $ E = m/\sqrt{2} $, where $ m $ is the half-gap width and charge $ Q = 1/4 $. Comparison with the spectrum in Fig.~\ref{fig:msqrt2} shows close agreement, and further numerics on scaling confirm that the corner modes lie at the expected energy for large systems (see Appendix \ref{app:scaling}).

\begin{figure}[tbp]
	\centering
	\includegraphics[trim={.0cm .0cm .0cm .0cm}, clip,width=.8\linewidth]{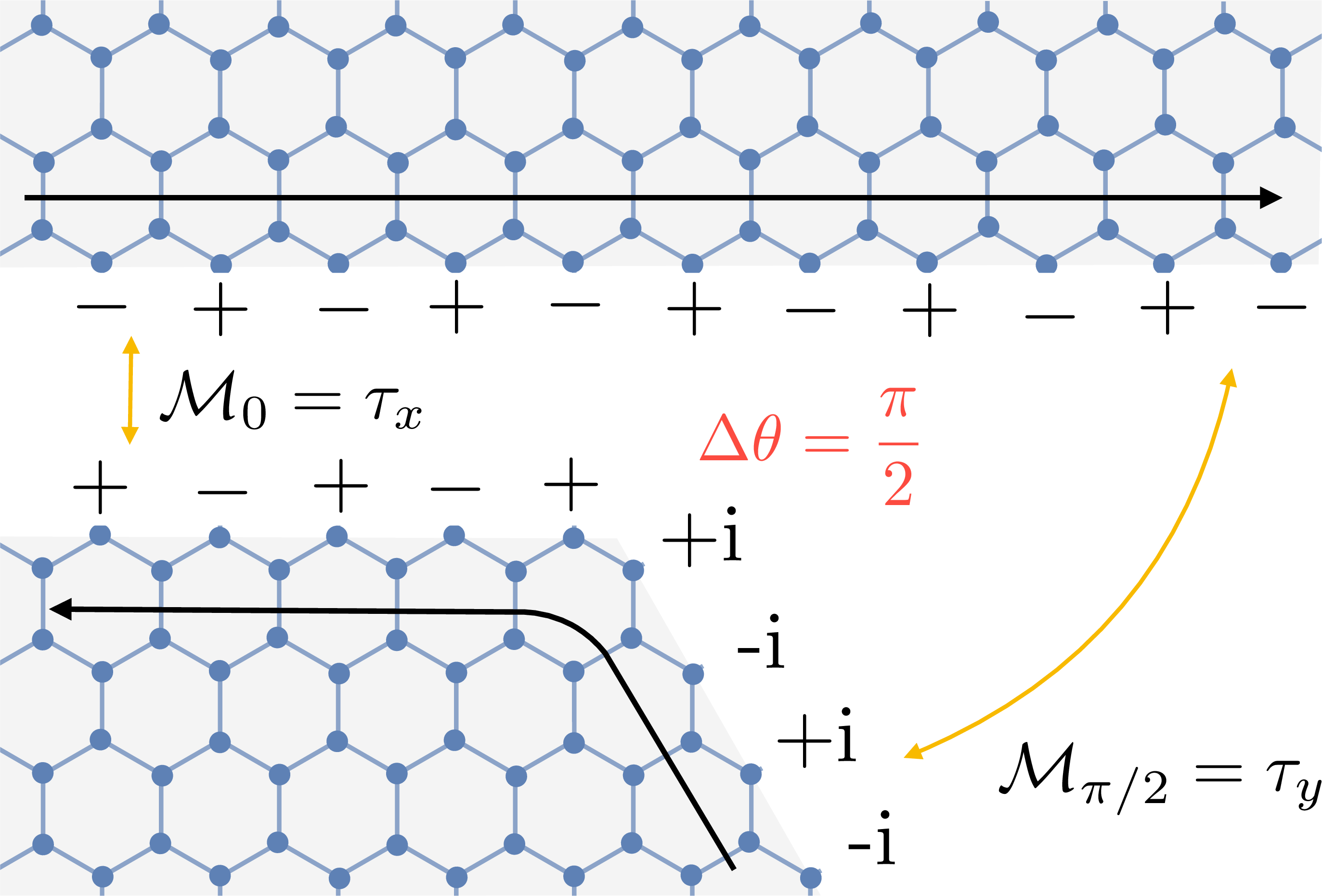}%
	\caption{\label{fig:gradisc}
		Construction of a $ \Omega = -2\pi/6 $ disclination in a Haldane model, demonstrating the connection between phase shifts of an edge state at a corner and disclination modes. This consists of two pieces, each having an edge state propagating counterclockwise, but one with a straight edge and therefore no phase shift and the other with a $ C_6 $ corner with a $ \pi/2 $ phase shift. When couplings are established between the pieces, masses will appear in the low-energy theory, gapping out the edge states. As the phase of the edge state shifts by $ \pi/2 $ at the corner, the masses will have the same $ \pi/2 $ shift, resulting in a localized mode at the disclination core with a charge of $ 1/4 $.
	}
\end{figure} 

\begin{figure}[tbp]
	\centering
	\includegraphics[trim={.0cm 0.0cm 0.0cm .0cm}, clip,width=.99\linewidth]{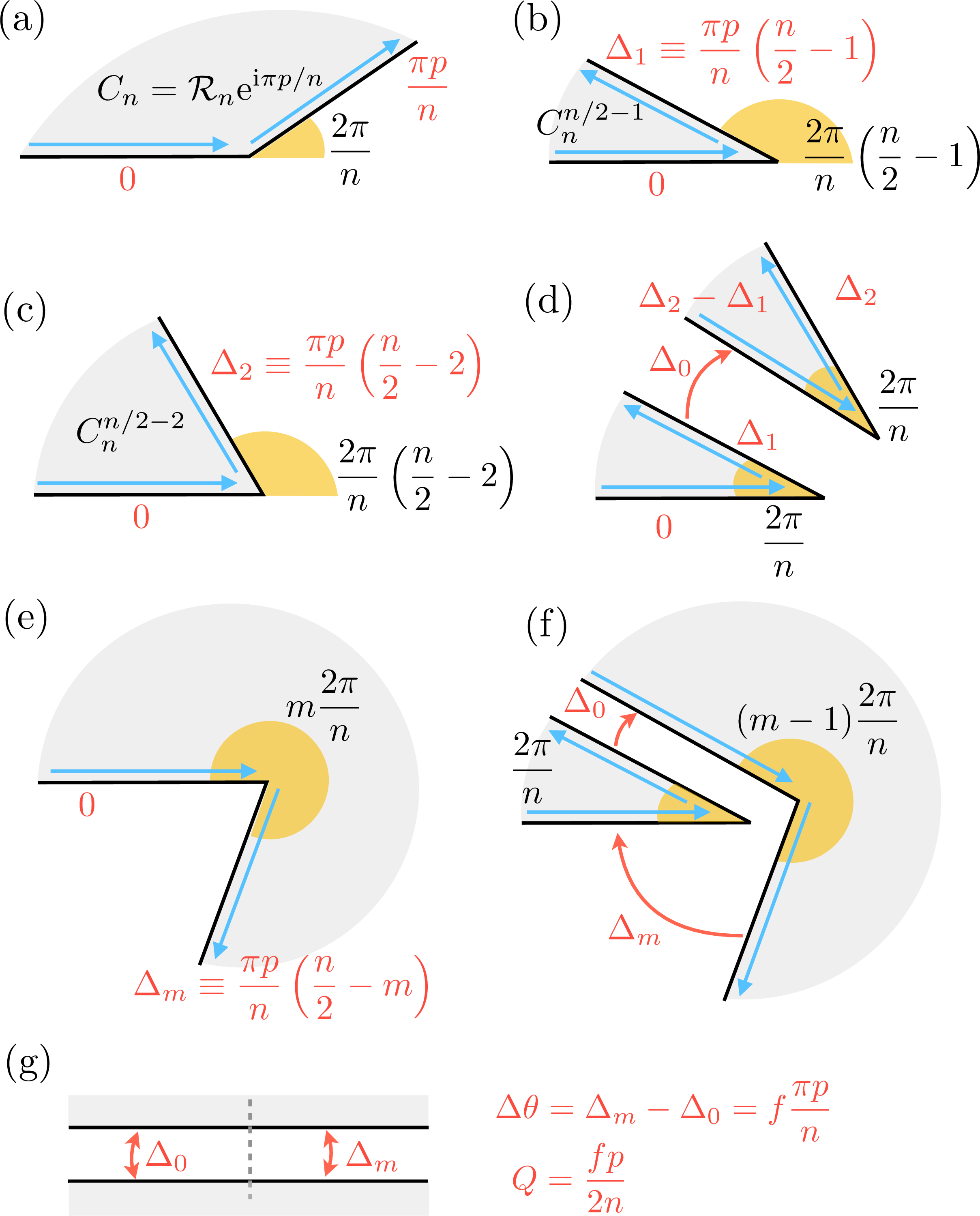}%
	\caption{\label{fig:disclin}
		Diagrams outlining the derivation of Eq.~\eqref{eq:discl} relating disclination charge to rotational symmetry $ n $, representation $ p $, and Frank index $ f $. (a) The phase $ \pi p/n $ of the $ C_n $ symmetry gives the phase acquired by an edge state as it passed a $ C_n $ corner. (b) The corresponding phase shift for an elementary corner of angle $ \pi-2\pi/n $ is given by the phase of $ C_n^{n/2-1} $. (c) Similarly, one may find the phase shift for a corner with twice this angle. (d) By splitting this piece into two, one finds a nonzero phase $ \Delta_0 $ across the cut in order for all phases to be consistent. We call this the gluing phase. (e) The phase shift at an arbitrary corner with angle $ 2\pi m /n $ is found by summing the gluing and elementary phases. (f) By splitting off an elementary piece, we may apply the same reasoning from Fig.~\ref{fig:gradisc} to find the change at a disclination with $ \Omega = -2\pi f/n = 2\pi(m-n)/n $.
	}
\end{figure}

\section{Disclinations\label{sec:disclinations}}

It is well known that the Haldane model (a single layer of the model we study) features localized modes with fractional charge at disclinations (rotational defects corresponding to the removal or addition of $ 2\pi/6 $ segments of the hexagonal lattice, forming conical points in three-dimensional space)~\cite{ruegg2013bound}. Given the presence of corner modes for a bilayer, this raises a possible connection between corner modes and disclination modes. In the following we show that there is indeed a connection between the two and derive the relationship
\begin{align}
Q_\text{disclination} = \frac{f}{2}Q_\text{corner} \label{eq:disclination}
\end{align}
between the disclination charge $ Q_\text{disclination} $ of a single Chern insulator layer and the corner charge $ Q_\text{corner} $ of a bilayer of Chern insulators with opposite Chern numbers.
Here $ f $ is the Frank index for the disclination angle (Frank angle) $ \Omega = -2\pi f/n $, with $ n $ being the rotational symmetry. A Frank index $ f > 0 $ ($ f < 0 $) corresponds to the removal (addition) of $ 2\pi/n $ segments; for example, the disclination shown in Fig.~\ref{fig:gradisc} has Frank index $ f = 1 $ and Frank angle $ \Omega = -2\pi/6 $.

Our approach to deriving \eqref{eq:disclination} uses the low-energy theory developed in the preceding section. The central idea is most clearly understood by first considering a specific example. Consider a $ \Omega = -2\pi/6 $ disclination in the Haldane model~\cite{ruegg2013bound}. One constructs this disclination by gluing two Haldane pieces, as shown in Fig.~\ref{fig:gradisc}. A piece with a flat edge is connected to a piece with a $ 2\pi/6 $ corner. As this corner is identical to the corner of a sixfold hexagonal sample, the phase shift for the edge state across this corner is $ \pi/2 $, as found in the preceding sections. When the two pieces are joined, the oppositely propagating edge states will gap out; however, because of this $ \pi/2 $ phase shift, the mass matrices will have a $ \dtheta = \pi/2 $ phase shift at the core of the disclination. Appealing to the theory in the preceding sections, one can immediately determine that a bound state will be present with charge $ 1/4 $.

We generalise this procedure to all disclination angles and rotational symmetries by deriving rules for the fractional phase shift at an ``elementary'' corner and the ``gluing'' phase for joining these pieces together, as shown in Fig.~\ref{fig:disclin}. We consider a two-dimensional $ n $-fold symmetric system with a single chiral edge state (Chern number of one). The edge theory must have a $ C_n $ symmetry satisfying $ C_n^n = -1 $, that is, $ C_n = \cR_{n} \e^{\ii \pi p/n} $, with odd $ p $ indexing a particular representation of this symmetry. If a disclination with Frank index $  f $ is made, one may expect a topologically bound mode at the disclination core depending on the representation $ p $. 

We carry this out for even rotational symmetries and then argue that the same result can be extended to odd rotational symmetries. 
We wish to find the phase acquired by the edge state at the most acute corner of angle $ 2\pi/n $; we refer to this as an elementary corner. The phase in $ C_n = \cR_{n} \e^{\ii \pi p/n} $ is associated with a $ 2\pi/n $ rotation of the state. At a corner with angle $ 2\pi/n $, the state is rotated by $ \pi - 2\pi/n = (2\pi/n)(n/2-1) $, which is $ n/2-1 $ times the rotation for $ C_n $. One therefore finds the phase for the elementary corner from $ C_n^{n/2-1} = \cR_{n/(n/2-1)}\e^{\ii(\pi p/n)(n/2-1)} $. We call this phase shift $ \Delta_1 \equiv (\pi p/n)(n/2-1) $. Similarly, we can find the phase shift of the second most acute corner of angle $ 4\pi/n $ is $ \Delta_2 \equiv (\pi p/n)(n/2-2) $. 

As shown in Fig.~\ref{fig:disclin}, if one splits this second smallest corner into two elementary pieces, in order for all phases to be consistent an additional gluing phase is needed between elementary pieces of $ \Delta_0 = -\pi p/n $. 
The phase shift $ \Delta_m $ across a ``corner'' with angle $ 2\pi m/n $ can then be found by summing over elementary and gluing phases, $ \Delta_m = m\Delta_1+(m-1)(-\pi p/n) = (\pi p/n) (n/2-m) $. The final step in finding the disclination charge is achieved by splitting the part consisting of $m$ pieces into two parts: one with $m-1$ pieces, and another with one piece. In joining these pieces, masses with phases $ \Delta_m $ and $ \Delta_0 $ will occur on either side of the disclination. 

Therefore, the phase difference between the two channels is $ \dtheta = \Delta_m - \Delta_0 = \pi p - m\pi p/n $. Since a sample consisting of $ m $ pieces will have Frank index $ f = n - m $, we have $ \dtheta = f\,\pi p/n $. The final expression for the disclination charge is therefore 
\begin{align}
Q = \frac{fp}{2n}. \label{eq:discl}
\end{align} 

For odd rotational symmetries the approach above does not directly apply. Nevertheless, one can expect the results to be consistent in the following sense. Consider disclinations with odd rotational symmetry $ n \in 2\mathbb{Z}+1 $. These disclinations can be found in a system with $ n' = 2n $ rotational symmetry by restricting the Frank index to even integers $ f' = 2f $. In this case, one therefore has disclination charge $ Q = f'p/2n' = (2f)p/2(2n) = fp/2n $, which is consistent with \eqref{eq:discl}. We therefore postulate that the result in \eqref{eq:discl} holds true for systems with odd rotational symmetry.

The final connection to corner charges of a bilayer is made by noting that a bilayer will have the following representation for $ n $-fold rotational symmetry $ C_n = \cR_{n}\exp(\ii \tau_z \pi p/n) $. From the preceding section, we know that this has corner charges $ Q = p/n $. Therefore, the corner charge for a bilayer is related to the disclination charge in a single layer by Eq.~\eqref{eq:disclination}. We remark that a similar result was obtained in Ref.~\cite{benalcazar2019quantisation} for a gapped single layer using different methods. Our result superficially differs by a factor of $ 1/2 $; nevertheless, the results are consistent. This is because our result relates a corner charge for a bilayer to a disclination charge in a single layer, whereas in Ref.~\cite{benalcazar2019quantisation} the corner charge of a single layer is related to the disclination charge of single layer.

\begin{figure*}[tbp]
	\centering
	\includegraphics[trim={.0cm 0cm 0.0cm .0cm}, clip,width=.75\linewidth]{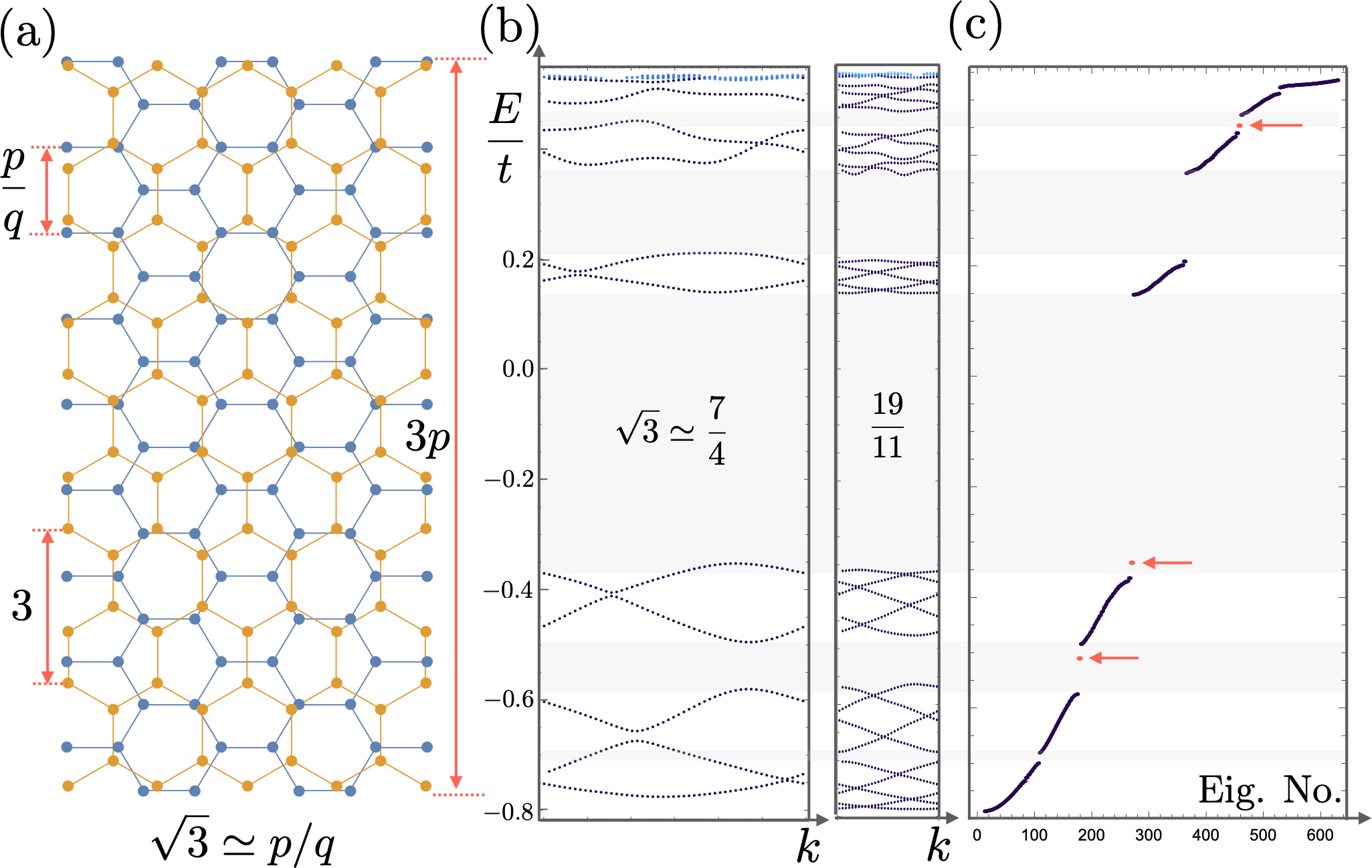}%
	\caption{\label{fig:C4edge} Hierarchy of edge gaps. (a) The ``strip'' geometry used to calculate the edge spectrum. An approximant unit cell is made by using the rational approximant $ \sqrt{3} \simeq p/q $, giving a unit cell length $ 3p $, dependent on the order $ p $ of the rational approximation. (b) The edge spectra for two rational approximations to $ \sqrt{3} $. One sees a folding occurring at higher-order approximants. This folding generically leads to anticrossings that can gap out under the presence of the interlayer coupling. (c) The spectrum of a sample with $ C_4 $ symmetry for $ \lambda_\perp = 5t $ and $ \lambda_\text{H} = 0.3t $. The corner states for the principal zero-energy gap are present, in addition to a number of distinct corner states within higher-order gaps.
	}
\end{figure*}

\section{Hierarchy of corner states\label{sec:hierarchy}}

In the preceding sections, all results relied solely on the presence of rotational symmetry. These were general and as such applied to both periodic and quasiperiodic systems.  
Nevertheless, quasiperiodicity itself can lead to a number of novel features. Here we discuss a striking example, demonstrable in the model we study. This is found by moving beyond the low-energy theory and instead looking at the full edge spectrum. As shown in Fig.~\ref{fig:C4edge}, for sufficiently large interlayer coupling and sample size, a hierarchy of additional gaps appears that can also harbor corner-localized modes.

The origin of these additional gaps is easily understood by attempting to construct the 1D spectrum of a strip geometry. Since the system is quasiperiodic, this cannot be exactly achieved; however, one can approach this by using ``approximants''. That is, we use the Diophantine approximations, $ \sqrt{3} \simeq p/q $, to produce a series of sequentially more accurate, larger unit-cell approximations to a true, infinite unit cell. Indeed, for a particular approximant, one can show that the unit-cell length is $ 3p $. The 1D Brillouin-zone width is therefore $ 2\pi/3p $ and reduces with approximant order. This leads to a folding back of the two edge-state bands.

At lowest order, the only crossing is at zero energy, resulting in the ``principal'' (largest) energy gap. However, at higher orders one has more crossings and, as such, more gaps. The edge spectrum plot in Fig.~\ref{fig:C4edge} demonstrates that gaps that appear at lower orders remain robust at higher orders. Moreover, these match well with the gaps found in the spectrum of a sample with $ C_4 $ symmetry. The corner-localized states in this spectrum are highlighted in red. Since each higher-order gap is formed from an anticrossing between the edge modes, the same low-energy theory as that used above must apply. However, it is an open question whether all higher-order gaps must share the same representation and therefore the same corner-state charge. 

For any finite sample the full hierarchy of gaps will not be resolved; as such, the number of corner states grows extensively with sample size. One can understand this in two ways. For an edge length $L$, one will not see $ k $-space features with wavelength larger than $L$, that is, momentum smaller than $ 1/L $; this provides a natural cutoff for relevant momentum transfers, providing a cutoff for relevant gaps. A similar cutoff can be obtained by considering the energy spacing for a sample of finite edge length $ L $. As in Eq.~\eqref{eq:edgequantisation}, the spacing is $ \sim1/L $; as such, all gaps smaller than this will not be resolved in the spectrum of the finite sample.

\section{Generalisations\label{sec:generalisations}}

\subsection{Stacking construction}

We highlight that while we have used a quasicrystalline model consisting of two incommensurately stacked crystalline layers, the same results apply to fully quasicrystalline lattices such as Penrose and Ammann-Beenker. For example, a bilayer consisting of coupled-opposite-Chern-number Penrose tilings~\cite{bandres16topological} would have a low-energy edge theory similar to that found here. 

Nevertheless, the stacking with a twist construction is a very natural and powerful way to produce similar phases. Indeed, doubling the rotational symmetry will generally lead to quasicrystalline rotational symmetries, that is, $ 4 $-fold $ \to 8 $-fold and $ 6 $-fold $ \to 12 $-fold. In addition the square root encountered in going from $ C_n \to C_{2n} $ can produce a nontrivial HOTI from a trivial HOTI. For example, a system with $ C_n = -\tau_0 $ after twisting will have $ C_{2n} = \tau_y K $, which protects $ Q=1/2 $ corner modes at the $ 2n $ corners. Also, as we find here, a system with conventional mass inversions without a twist will have fractional mass kinks with a twist. 

\subsection{Interlayer coupling}

The model studied throughout this work had an interlayer coupling which differed from the interlayer Rashba coupling in the Kane-Mele model. A natural question is whether this is important to the results described throughout. The answer is no; we could, indeed, use a generalized Rashba type of interlayer coupling 
\begin{align}
H_\text{R} = \ii \lambda_\text{R} \sum_{ij} t^\perp_{ij} c_i^\dagger \left( \ttau \times \bm{d}_{ij} \right)_z c_i,
\end{align}
where $ \bm{d}_{ij} = \rr_i - \rr_j $. Due to this different interlayer coupling, the $ C_{12} $ symmetry is changed from $ \cR_{12}\tau_x K $ to $ \cR_{12} \ii\tau_y K $. That is, the antiunitary part is a spinful time-reversal symmetry $ (\ii \tau_y K)^2 = -1 $ opposed to spinless $ (\tau_x K)^2 = 1 $.  However, all of our numerical results remain qualitatively unchanged, highlighting that the time-reversal part of the $ C_{12} $ symmetry is irrelevant to the protection of the HOTI phase. The important feature is the nonlocal action of the symmetry.

\section{Summary}

We have shown that a simple \ang{30} twist of the Kane-Mele model is sufficient to produce a quasicrystalline higher-order topological insulator. We showed this result numerically and provided a detailed analytical understanding based on a low-energy theory. In carrying this out, we derived a general low-energy theory that classifies higher-order topological phases with any rotational symmetry, including those disallowed in periodic crystalline materials. We found that these are generally associated with what we dub fractional mass kinks, in which instead of a change in sign in the low-energy mass, there is a fractional shift in phase. We then highlighted a natural connection between corner modes and disclination modes and used the low-energy theory to establish this relationship in general. Furthermore, we demonstrated numerically that for strong interlayer couplings, a hierarchy of gaps opens in the edge spectrum. We showed this to be a direct result of the quasiperiodicity. Finally, we outlined a number of natural generalizations and highlighted that our stacking with a twist construction can, in general, produce a nontrivial higher-order topological insulator from a trivial system without a twist.

\section{Acknowledgments}

We gratefully acknowledge support from EPSRC via Grant No. EP/P034616/1 and from a Simons Investigator Award of the Simons Foundation.

\appendix

\section{Derivation of edge theory \label{app:derivation}}

We start with the bulk low-energy theory for a bilayer Haldane model
\begin{align}
\cH_\text{bulk} = \left( k_x\sigma_x\rho_z + k_y\sigma_y \right) + \dH \sigma_z \rho_z \tau_z + \dS \sigma_z, \label{lowenergybulk}
\end{align}
where $ \sigma $ is the sublattice (A or B) degree of freedom, $ \rho $ is the valley ($ K $ or $ K' $), and $ \tau $ is the layer (top or bottom). The first two terms derive from the nearest-neighbor hopping and describe the two Dirac cones. The term proportional to $ \dH $ derives from the Haldane time-reversal-symmetry breaking  next-nearest-neighbour hopping. The term proportional to $ \dS $ is due to a sublattice-dependent on-site energy and is included to allow for a normal insulator phase for $ \dS > \dH > 0 $. We will assume throughout that the interlayer hopping $ \lambda_\perp = 0 $, adding this back in at the end. Note that this derivation is identical to that of the Kane-Mele model by associating layer and spin degrees of freedom.

Our first step will be to perform a unitary rotation in the $ \sigma $-$\rho $ subspace to remove the opposite chirality between valleys,
\begin{align}
\cH_\text{bulk}' \equiv U \cH_\text{bulk} U = \kk \cdot \ssigma + \dH \sigma_z\tau_z + \dS \sigma_z\rho_z,
\end{align}
with 
\begin{align}
U \equiv \begin{pmatrix}
\sigma_0 & 0\\
0 & \sigma_y
\end{pmatrix} \tau_0. \label{eq:edgeU}
\end{align}

Now consider a domain wall with normal $ \hnn $ and tangent $ \htt = \hzz \times \hnn $ between  nontrivial and trivial regions. We decompose $ \kk = k_t \htt + k_n \hnn $. As the Hamiltonian is now position dependent along the $ \hnn $ direction we write $ k_n = -\ii\partial_\lambda $, where $ \lambda $ is the distance from the domain wall. Letting $ k_t = k $, we have
\begin{align}
\cH_\text{bulk}' = k \htt \cdot \ssigma - \ii \partial_\lambda \hnn\cdot\ssigma + \dH \sigma_z\tau_z +\dS\sigma_z\rho_z,
\end{align}
and use a mass dependence~\cite{cayssol2013various}
\begin{align}
&\dH > 0, \ \dS =0 \quad \text{for} \quad \lambda<0,\\
&\dH = 0, \ \dS >0 \quad \text{for} \quad \lambda>0.
\end{align}
Therefore, for $ \lambda < 0 $, we have Chern numbers $ C = \pm 1 $ on each layer, while for $ \lambda >0  $ both layers have $ C = 0 $.

We then search for a solution exponentially localized at the domain wall,
\begin{align}
\Psi(\lambda,k) = \left\{
\begin{array}{ll}
\e^{\dH\lambda}\, \psi(k), \quad \quad &\lambda <0, \\
\e^{-\dS\lambda}\, \psi(k), \quad \quad &\lambda >0.
\end{array}
\right.
\end{align}
Substituting into $ \cH_\text{bulk}\Psi = E\Psi  $,
\begin{align}
&\left[ k\htt\cdot\ssigma - \ii\dH\hnn\cdot\ssigma + \dH\sigma_z\tau_z \right] \Psi = E\Psi, \ &\lambda<0\\
&\left[ k\htt\cdot\ssigma + \ii\dS\hnn\cdot\ssigma + \dS\sigma_z\rho_z \right] \Psi = E\Psi, \ &\lambda>0.
\end{align}

This can be solved by requiring the following,
\begin{align}
\dH &\left[ - \ii\hnn\cdot\ssigma + \sigma_z\tau_z \right] \psi = 0,\nonumber\\
\dS &\left[  +\ii\hnn\cdot\ssigma + \sigma_z\rho_z \right] \psi = 0,\label{eq:conditions}
\end{align}
and,
\begin{align}
\underbrace{(k \htt\cdot\ssigma)}_{\cH_\text{edge}}\psi = E\psi.
\end{align}
One can rewrite the first two conditions in terms of projections. Using $ \ii \hnn\cdot\ssigma = \sigma_z(\hzz\times\hnn)\cdot\sigma = \sigma_z \htt\cdot\ssigma $, one has
\begin{align}
\sigma_z\tau_z\left[ 1-\htt\cdot\ssigma\tau_z \right]\psi = 0,\\
\sigma_z\rho_z\left[ 1+\htt\cdot\ssigma\rho_z \right]\psi = 0,
\end{align}
which can be written as
\begin{align}
\frac{1}{2}(1+\htt\cdot\ssigma\tau_z)\psi = \psi,\\
\frac{1}{2}(1-\htt\cdot\ssigma\rho_z)\psi = \psi,
\end{align}
where we define
\begin{align}
P_\text{H} \equiv \frac{1}{2}(1+\htt\cdot\ssigma\tau_z),\\
P_\text{S} \equiv \frac{1}{2}(1-\htt\cdot\ssigma\rho_z),
\end{align}
satisfying $ P_\text{H}^2 = P_\text{H} $, $ P_\text{S}^2 = P_\text{S} $, and $ \left[P_\text{H},P_\text{S}\right] = 0 $.
The above projectors imply that the internal structure of the edge theory is lower dimensional than that of the bulk theory. In order to find this lower-dimensional subspace we find rotations that diagonalize the composite projector $ P \equiv P_\text{H} P_\text{s} $.

\begin{figure}[tbp]
	\centering
	\includegraphics[trim={.0cm .0cm .0cm .0cm}, clip,width=1.\linewidth]{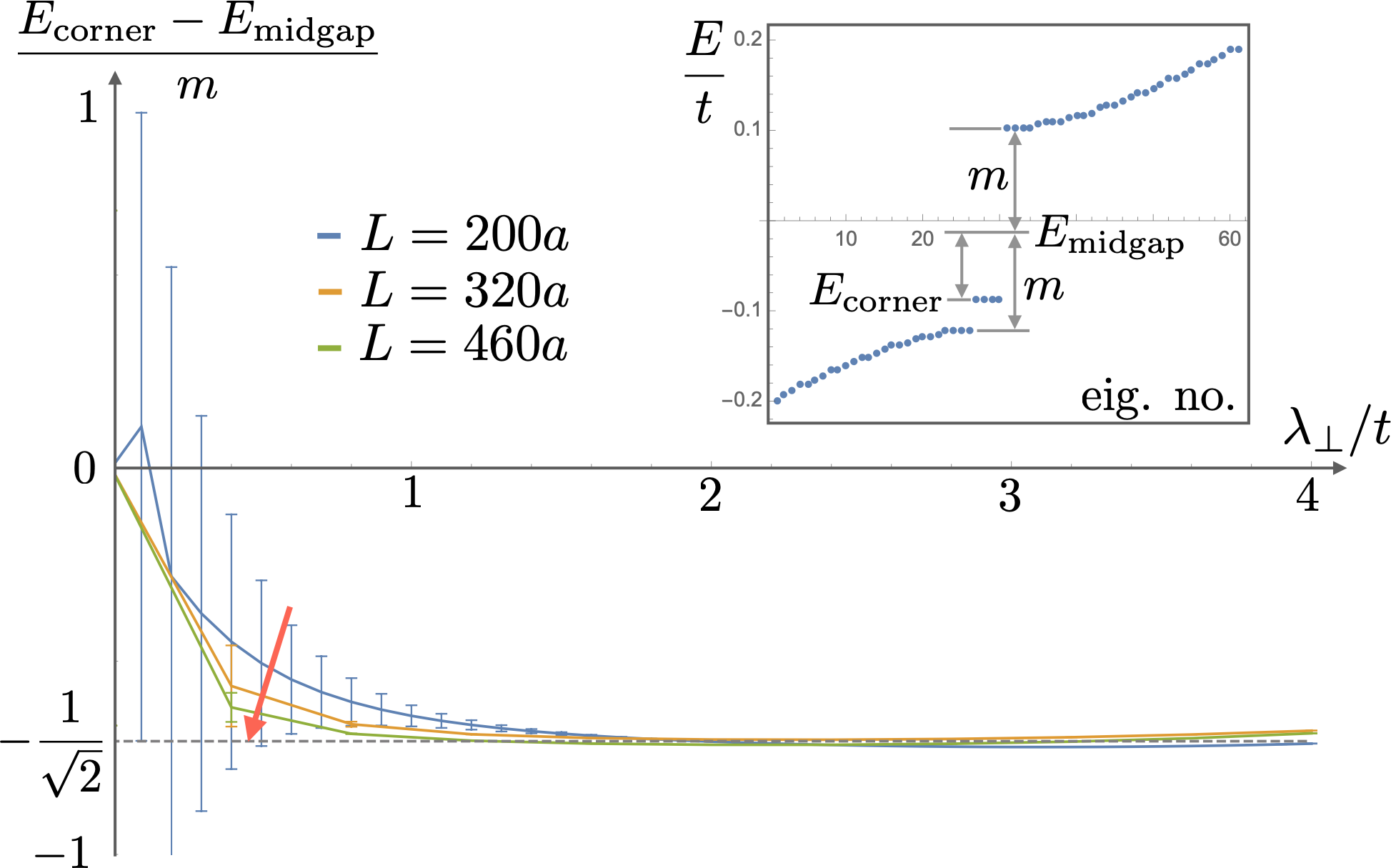}
	\caption{\label{fig:msqrt2scaling} The in-gap location of the corner states for a $ C_4 $ symmetric system as a function of interlayer coupling $ \lambda_\perp $ and for various system sizes. The expected location $ E/m = -1/\sqrt{2} $ is indicated by the dashed line. The error bars indicate the splitting of the four degenerate corner states. This splitting is due to a lengthening of the exponential tails of the corner states for small $ \lambda_\perp $. (Inset) Diagram indicating how the various parameters are defined for the numerically obtained spectrum.
	}
\end{figure}

The composite projector $ P \equiv P_\text{H} P_\text{s} $, satisfying $ P \psi = \psi $, can be alternatively decomposed as
\begin{align}
P &= \frac{1}{2}(1+\htt\cdot\ssigma\tau_z)\frac{1}{2}(1-\htt\cdot\ssigma\rho_z),\\
&= \frac{1}{4}(1-\tau_z\rho_z + \htt\cdot\ssigma\tau_z - \htt\cdot\ssigma\rho_z),\\
&=\underbrace{\frac{1}{2}(1+\htt\cdot\ssigma\tau_z)}_{\equiv P_1}\underbrace{\frac{1}{2}(1-\tau_z\rho_z)}_{\equiv P_2}.
\end{align}
One can now diagonalize $ P_1 $ and $ P_2 $ using the following unitary transformations
\begin{align}
&V_1 \equiv \exp(\ii \pi \hnn\cdot\ssigma \tau_z) , \quad V_1 P_1 V_1^\dagger = \frac{1}{2}(1-\sigma_z)\rho_0\tau_0,\nonumber\\
&V_2 \equiv \sigma_0 \begin{pmatrix}
\rho_0 & 0 \\
0 & \rho_x
\end{pmatrix}, \quad V_2 P_2 V_2^\dagger = \frac{1}{2}\sigma_0(1-\rho_z)_0\tau_0,\label{eq:edgeV}
\end{align}
with $ \left[ V_1,V_2 \right] = 0 $, $ \left[ V_1,P_2 \right] = 0 $, and $ \left[ V_2,P_1 \right] = 0 $.
By noting that the projector
\begin{align}
\frac{1}{2}(1-\sigma_z) = \begin{pmatrix}
0 & 0\\
0 & 1
\end{pmatrix}
\end{align}
picks out the ``down'' subspace, one can construct the rectangular matrix
\begin{align}
p \equiv \begin{pmatrix}
0 & 0 & 0 & \tau_0
\end{pmatrix}^\text{T}, \label{eq:edgep}
\end{align}
which picks out the correct subspace for $ P' $.

One therefore finds the final edge Hamiltonian by taking this projection:
\begin{align}
\left(p^\text{T} V \cH_\text{edge} V^\dagger p \right) \left(p^\text{T}V\psi\right) &= E \left(p^\text{T}V\psi\right),\\
\cH'_\text{edge} \psi' &= E \psi',
\end{align}
where $ V \equiv V_2 V_1 $ and
\begin{align}
\cH'_\text{edge} &= p^\text{T} V \left( k \htt\cdot\ssigma  \right) V^\dagger p\\
&=k\tau_z. \label{eq:edgefinal}
\end{align}
The final result is a one-dimensional Dirac theory, with two counterpropagating modes. Note that due to the rotation $ V $ that mixes all three degrees of freedom, the final $ \tau $ cannot be directly associated with the original layer degree of freedom.

In summary one has
\begin{align}
\cH'_\text{edge} &= \left(p^\text{T} V U \right) \cH_\text{bulk} \left(U^\dagger V^\dagger p \right)\\
&= k\tau_z,
\end{align}
with $ U $, $ V $, and $ p $ defined in \eqref{eq:edgeU}, \eqref{eq:edgeV}, and \eqref{eq:edgep}. To connect back to the main text, we reiterate that the interlayer coupling was set to zero throughout this derivation; therefore, this is the edge theory for two uncoupled layers. In order to incorporate the interlayer coupling, one notes (as in the main text) that since numerically $ \lambda_\perp \neq 0 $ gaps out the edge theory, one must include all terms that can gap out \eqref{eq:edgefinal}, that is, $ \tau_x  $ and $ \tau_y $.

\section{Scaling of the corner-state in-gap energy \label{app:scaling}}

In the main text it was found that the corner-localized states for a system with 12-fold symmetry will have an energy of $ E = m/\sqrt{2} $ above the midgap value $ E_\text{midgap} $, where $ m = E_\text{gap}/2 $ is the half-gap width. Similarly, by using $ C_4 = C_{12}^3 $, one can show that for a system with fourfold symmetry the corner states will similarly have an energy of $ E = -m/\sqrt{2} $.

In Fig. \ref{fig:msqrt2scaling} we plot the numerically obtained in-gap energy of the corner-localized states of a system with fourfold symmetry as a function of interlayer coupling and for various system sizes. For $ \lambda_\perp/t \gtrsim 1 $ the agreement is good; however, this does begin to deviate for larger $ \lambda_\perp $. This is expected since the result $ E = -m/\sqrt{2} $ is found from a low-energy theory and should therefore be expected to remain valid only for moderate couplings. One also notices a strong deviation for $ \lambda_\perp \lesssim 1 $. Again, this is expected since for a finite system size of length $ L $, there is a natural energy scale $ v_F/L $ discussed in Sec. \ref{sec:hierarchy} and evident in Eq. \eqref{eq:edgequantisation}, which sets a resolution on spectral features. For increasing system size, this resolution becomes sharper, and accordingly the curves in Fig. \ref{fig:msqrt2scaling} flatten towards $ E = -m/\sqrt{2} $.


%

\end{document}